\documentclass[12pt]{article}

\begin{document}
\input {epsf}

\newcommand{\beq}{\begin{equation}}
\newcommand{\eeq}{\end{equation}}
\newcommand{\beqa}{\begin{eqnarray}}
\newcommand{\eeqa}{\end{eqnarray}}

\def\ov{\overline}
\def\onlyif{\rightarrow}

\def\openone{\leavevmode\hbox{\small1\kern-3.8pt\normalsize1}}

\def\a{\alpha}
\def\b{\beta}
\def\g{\gamma}
\def\r{\rho}
\def\minus{\,-\,}
\def\eks{\bf x}
\def\kay{\bf k}

\def\ket#1{|\,#1\,\rangle}
\def\bra#1{\langle\, #1\,|}
\def\braket#1#2{\langle\, #1\,|\,#2\,\rangle}
\def\proj#1#2{\ket{#1}\bra{#2}}
\def\expect#1{\langle\, #1\, \rangle}
\def\trialexpect#1{\expect#1_{\rm trial}}
\def\ensemblexpect#1{\expect#1_{\rm ensemble}}
\def\kpsi{\ket{\psi}}
\def\kphi{\ket{\phi}}
\def\bpsi{\bra{\psi}}
\def\bphi{\bra{\phi}}

\def\ditto{\rule[0.5ex]{2cm}{.4pt}\enspace}
\def\th{\thinspace}
\def\ni{\noindent}
\def\thirty{\hbox to \hsize{\hfill\rule[5pt]{2.5cm}{0.5pt}\hfill}}

\def\set#1{\{ #1\}}
\def\setbuilder#1#2{\{ #1:\; #2\}}
\def\Prob#1{{\rm Prob}(#1)}
\def\pair#1#2{\langle #1,#2\rangle}
\def\Id{\bf 1}

\def\dee#1#2{\frac{\partial #1}{\partial #2}}
\def\deetwo#1#2{\frac{\partial\,^2 #1}{\partial #2^2}}
\def\deethree#1#2{\frac{\partial\,^3 #1}{\partial #2^3}}

\newcommand{\xx}{{\scriptstyle -}\hspace{-.5pt}x}
\newcommand{\yy}{{\scriptstyle -}\hspace{-.5pt}y}
\newcommand{\zz}{{\scriptstyle -}\hspace{-.5pt}z}
\newcommand{\kk}{{\scriptstyle -}\hspace{-.5pt}k}
\newcommand{\sx}{{\scriptscriptstyle -}\hspace{-.5pt}x}
\newcommand{\sy}{{\scriptscriptstyle -}\hspace{-.5pt}y}
\newcommand{\sz}{{\scriptscriptstyle -}\hspace{-.5pt}z}
\newcommand{\sk}{{\scriptscriptstyle -}\hspace{-.5pt}k}

\def\openone{\leavevmode\hbox{\small1\kern-3.8pt\normalsize1}}

\title{Quantum key distribution with trusted quantum relay}
\author{H. Bechmann-Pasquinucci$ ^{1,2}$, A. Pasquinucci$ ^{2}$
\\
\small
$^{1} ${\it  University of Pavia,  Dipartimento di Fisica
``A.\ Volta", via Bassi 6, I-27100 Pavia, Italy}\\ 
\small
$^{2} ${\it {\rm UCCI.IT}, via Olmo 26, I-23888 Rovagnate (LC), Italy 
}}
\maketitle

\abstract{A trusted quantum relay is introduced to enable quantum key 
distribution links to form the basic legs in a quantum 
key distribution network. The idea is based on the well-known 
intercept/resend eavesdropping. The same scheme can be used to make 
quantum key distribution between several parties. No entanglement is 
required.} 
\vspace{1 cm} 
\normalsize

\section{Introduction}
In the field of quantum information, quantum key distribution 
\cite{BB84,geneva} is the 
application which is more developed, to the point that commercial 
prototypes exist.
The strength of quantum key distribution namely that security is 
based on the basic laws of quantum physics, is somehow also its weakness. 
The 
fact that measurements will disturb the state, that perfect quantum 
cloning is not possible, makes it impossible to make, for example, perfect 
quantum repeaters, which puts a limitation on the practical 
implementations.

Until now quantum key distribution has mainly been considered a 
point-to-point link between Alice and Bob.
Howover, recently discussion has started on how to form quantum key
distribution networks. And not only theoretical networks where one can
imagine using distributed multipartite entanglement and prolong the
distance with entanglement swapping, but practical networks which can be
implemented with todays technology.

We suggest a very simple quantum relay, which when trusted can be used as 
a basic building block in forming a network. The relay is basically 
performing the well-known intercept/resend eavesdropping strategy 
\cite{expcryp}, but is 
cooperating with Alice and Bob. The protocol can also be seen as a 
concatenation of quantum key distribution protocols. The requirement is 
that the relay has to 
be trusted, since in principle the relay will know the key generated by 
Alice and Bob.

As always nothing comes for free, and the price that Alice and Bob 
have to pay is a lower key generation rate. 
This may put some practical limitations to how many relays a network 
can contain. On the other hand, due to the simple working structure of the 
quantum relay, it is easy to implement different parts of a network with 
different quantum key distribution platforms. Which means that some parts 
of the network can be carried out by fiber implementations, for example 
within cites, whereas the connection between cities could be carried out 
by, for example, free space implementation either by line of sight or even 
ground to satellite.

It should be stressed that as far as a QKD network is built without
quantum repeaters using entanglement swapping, teleportation etc.\ 
and is implemented only
in fibers and with the quantum relays we introduce in this paper, 
the maximum distance is still limited to around 100km. This point will
be discussed in details in section 5.

We base the protocol on the BB84 protocol for quantum cryptography 
\cite{BB84}. 
However, it should be possible to consider a similar scenario using other 
protocols for quantum key distribution.

The paper is organized as follows: In section 2, we present the simplest 
scenario, Alice and Bob and in the middle a trusted quantum relay, Trent 
\cite{BS}. 
In section 3, we introduce the eavesdropper and consider simple 
intercept/resend eavesdropping. 
Section 4 is dedicated to a discussion of 
how to use the proposed protocol for multi partite quantum key distribution 
and of how to get the most out of the data. 
Section 5 is dedicated to a discussion of the impossibility of extending 
the distance between Alice and Bob in some situations.
In section 6, we discuss 
the network structure and topology of a simple network for quantum 
key distribution. 
Finally, in section 7 we conclude.

\section{Trusted quantum relay: the protocol}
\begin{figure}[t]
\begin{center}
\begin{picture}(100,30)(0,0)
\put(0,5){\framebox(20,15){X,Y}}
\put(40,5){\framebox(20,15){X,Y}}
\put(80,5){\framebox(20,15){X,Y}}
\put(5,0){Alice}
\put(45,0){Trent}
\put(85,0){Bob}
\put(20,12.5){\line(1,0){20}}
\put(60,12.5){\line(1,0){20}}
\end{picture}
\caption{The basic scenario: Alice and Bob are connected via the trusted 
quantum relay station, Trent.}
\end{center}
\end{figure}
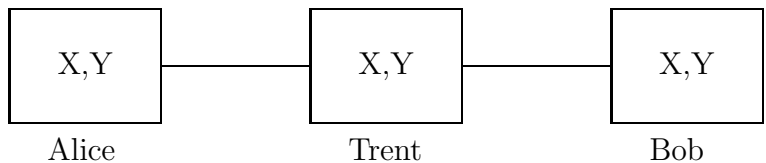
Suppose that Alice and Bob want to establish a secret key by means of 
quantum key distribution and that they use the BB84 protocol, however 
their conditions are such that the need to go via a trusted quantum relay, 
Trent. Trent's role in the protocol is basically to perform the well-known
intercept/resend eavesdropping --- but afterwords assist Alice and Bob in 
the sifting procedure. Step by step the protocol becomes:
\begin{enumerate}
\item Alice prepares a qubit in one of the four states $\pm x$ or $\pm 
y$, and sends it to Trent
\item Trent measures, as if he was Bob, in either the $X$ or the $Y$ 
basis. According to the result of his measurement, he prepares the same 
state that he found in this measurement and sends it to Bob.
\item Bob measures in the $X$ or the $Y$ basis.
\item Alice, Trent and Bob repeat the first 3 steps many times.
\item {\sl The relay sifting procedure}: Trent announces in which basis has
measured each qubit, Alice and Bob take note of this; if there are more 
than 
one relay, each one of them announces the basis used for each qubit;
The role of Trent (or of all relays) is now over.
\item Alice and Bob sifting procedure: Alice and Bob announce to each other
which basis they used, and they keep only the qubits for which Alice, 
Trent 
and Bob used the same basis.
\item Alice and Bob proceed with the estimate of the error rate 
followed by error correction and privacy 
amplification to obtain a secret key --- as in the standard BB84 protocol.
\end{enumerate}
Notice that after step 5, Alice and Bob can use a different classical 
communication channel since Trent does not need to be informed of the 
following steps.
It is also clear that Alice and Bob loose more data during the sifting 
procedure than they do in the standard BB84 protocol. In a point to point 
link, they keep about $1/2$ of the raw data. Whereas with Trent in the middle 
there is only probability $1/4$ that they all used the same basis, which 
means that they can only keep $1/4$ of the data. In the presence of $N$
relays, this becomes $1/2^{N+1}$.

If Trent is acting correctly, that is he performs the 
measurements defined by the protocol, resends the states he finds and 
announces the bases he has used, then Alice and Bob theoretically should 
find no errors in the sifted key. The situation is hereafter the same as 
for the standard BB84; Alice and Bob can continue the classical part of 
the protocol and obtain a secret key.

It is clear that Alice and Bob need to trust Trent, since they (in 
ideal situations) all share the same raw data, hence if Trent listens
on the classical communication between Alice and Bob, he can obtain the same 
secret key. From this point of view, making Trent listen to the 
classical communication between Alice and Bob, leads to all three (or $N$)
of them to share the same secret key. This is correct if there will not be 
errors or attacks by Eve, as we will see in the next section.

\section{Including the eavesdropper, Eve}
As always the security is based on what Eve can do and 
how much information she can retrieve from the system. In this case the 
full analysis becomes more complicated because Eve can eavesdrop on two 
channels: Channel 1, from Alice to Trent and channel 2, from Trent to Bob.  

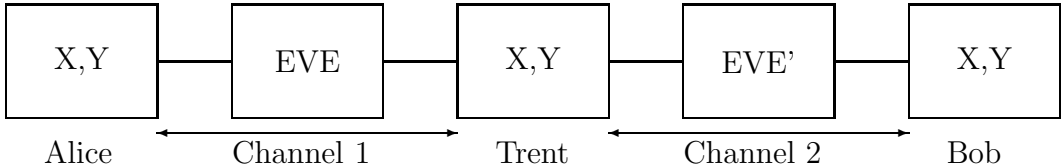
\begin{figure}[t]
\begin{center}
\begin{picture}(140,30)(0,0)
\put(0,6){\framebox(20,15){X,Y}}
\put(30,6){\framebox(20,15){EVE}}
\put(60,6){\framebox(20,15){X,Y}}
\put(90,6){\framebox(20,15){EVE'}}
\put(120,6){\framebox(20,15){X,Y}}
\put(5,0){Alice}
\put(65,0){Trent}
\put(125,0){Bob}
\put(20,4){\vector(1,0){40}}
\put(60,4){\vector(-1,0){40}}
\put(30,0){Channel 1}
\put(80,4){\vector(1,0){40}}
\put(120,4){\vector(-1,0){40}}
\put(90,0){Channel 2}
\put(20,13.5){\line(1,0){10}}
\put(50,13.5){\line(1,0){10}}
\put(80,13.5){\line(1,0){10}}
\put(110,13.5){\line(1,0){10}}
\end{picture}
\caption{The basic eavesdropping scenario: Alice-Eve-Trent-Eve' and Bob.}
\end{center}
\end{figure}

However, there is one very important point which should be kept in mind, 
namely that the two channels are used at different times. Which means that 
Eve at no point has access to both channels at the same time. This prevents 
Eve from doing some kinds of joint attacks on the two channels, at most 
she could let the same ancilla interact with both channels and make one 
measurement. However these general considerations are beyond the scope of 
this paper.
Here we will only consider very simple attacks like the  
intercept/resend eavesdropping in both 
channels, as these will give a first indication of the security of the 
protocol. Moreover, it has recently been shown 
\cite{helle}, that if the eavesdropper has no quantum 
memory, the optimal eavesdropping attack is actually the intercept/resend. 

Lets first consider Eve doing an intercept/resend attack. If she attacks 
only
one channel, either 1 or 2, there is no difference in the analysis with 
respect to the standard BB84. For example, the amount of errors she 
introduces in the
sifted key are the same, approximately 25\%. 

Consider then the case in which Eve attacks both channels with
intercept/resend. The first question is about the amount of information
that she can get by attacking both channels. Doing intercept/resend on
the first channel, for half of the qubits sent by Alice, Eve uses the
same basis, thus she obtains 1 bit of information per qubit, i.e. all the
information sent by Alice. For these qubits Eve can gain no more
information by eavesdropping on the second channel. For the other half
of the qubits sent by Alice, Eve uses the wrong basis on the first
channel, and obtains a fully random result, so 0 bit of information. 
Moreover she sends to Trent a random bit, which means that the result of
Trent is random, and  even if Eve would eavesdrop a second time on
the qubit sent by Trent, she would obtain a random result. Thus Eve
does not increase her information by eavesdropping on both channels.

Lets now consider the errors she introduces by eavesdropping twice, this
analysis will anyway give us some insights in the properties of our protocol.

If she attacks both channels with intercept/resend, she has to decide if 
she chooses at random the bases she uses for both channels, or in the second 
channel she uses the same basis she has used for the first. 
It is easy to see that if she chooses independently the bases in the 
two channels, she introduces approximately 37.5\% errors in Bob's sifted key.
This result can be obtained as follows. After she eavesdrops on the first 
channel, Eve introduces 25\% of error in Trent's sifted key. Since Eve
eavesdrops independently in the second channel, she will introduce 25\% of
errors on the 75\% of right qubits of the sifted key sent by Trent, 
thus adding other 18.75\% of wrong bits to Bob's sifted key. 
For the 25\% of wrong bits of the sifted key sent by Trent to Bob, 
again Eve introduces 25\% of errors for Bob with respect to Trent, 
but now this means that Bob would measure the correct bit with respect to 
Alice. So for 6.25\% of wrong bits of the sifted key sent by Trent to Bob,
Bob will measure a correct bit with respect to Alice. The total error
in Bob's sifted key is then $(25+18.75-6.25)=37.5\%$

Notice that Trent and Bob sifted keys are different, in particular among the
different bits there are 6.25\% of errors in Trent sifted key 
with respect to Alice sifted key, 
which are not present in Bob 
sifted key. Thus, when Alice and Bob run the error correction algorithm,
they will not correct these errors. Even if Trent follows Alice and Bob
error correction, Trent's key will still be different from Alice
and Bob's key.

Since we have assumed that Eve does independent attacks on both
channels, it follows that this result can be applied also to
experimental errors. Let assume that both channel 1 and 2 have 
independent
experimental errors, that is QBER value $D_i$ ($0\leq D_i \leq 1$). Thus
it follows that in Trent sifted key statistically there are $D_1 D_2$
errors not present in Bob's sifted key, with respect to Alice sifted key.
If Trent follows Alice and Bob error correction and privacy
amplification, he would get a key very similar but not identical to the
one shared by Alice and Bob. Notice that the knowledge of Trent on the
sifted key is very large ($D_i$ must be small for QKD to be able to
produce a secret key), thus it is not possible in general to
arbitrarily reduce Trent's information on the final key as it is done for
Eve in the privacy amplification phase of the protocol. Thus even if
Trent secret key will not be exactly identical to the one of Alice and
Bob, Trent will have too much information on the key and will have to be
trusted.

Alternatively Eve can use in the second channel the same basis she used
in the first. The analysis is similar to the previous one, and one 
obtains that Eve introduces in Bob sifted key 25\% of errors, as if she 
would eavesdrop just on one channel. But this time the number of wrong 
bits in Trent sifted key which are correct in Bob is up to 12.5\%. This 
different result is due to the fact that the errors introduced by Eve in 
the two channels are correlated and not independent. 

Notice that as long as Trent does not take part in the error correction, 
but only in the sifting procedure, it is impossible for Alice and Bob to 
know where an error has occurred, if it has occurred in the first or the 
second channel; that is effectively Alice and Bob  have only one long 
communication channel.

Even for optimal eavesdropping strategies, i.e.\ when Eve uses both an
ancilla and a quantum memory, we expect to find that the best that she can
do is to eavesdrop in only one channel. Except perhaps for the very
special situation where Eve let's the same ancilla interact with the qubit
in both channels. However, an analysis of these attacks is beyond the
scope of this paper.

\section{Key distribution between several parties}
\begin{table}
\begin{center}
\begin{tabular}{|l|c|c|c|c|c|c|c|c|}\hline
&1& 2& 3& 4& 5& 6& 7 & 8 \\ \hline 
Alice (A)& {\bf X} &{\bf X}  &X  &X          
&{\bf Y} &{\bf Y} & Y &Y \\ 
\hline
Carol (C)& {\bf X} &{\bf X}  &Y &{\bf Y} 
&{\bf Y} &{\bf Y} &X  &{\bf X} \\ 
\hline
Bob   (B)& {\bf X} &Y       &X  &{\bf Y} 
&{\bf Y} &X  &Y &{\bf X} \\ \hline
Secret Key & A-C-B& A-C& -- & C-B & A-C-B& A-C& --& C-B\\ \hline
\end{tabular}
\caption{The table displays all the different measurement situations, and 
how they can be used to create secret key 
between the different parties. Boldface indicates the parties which can 
generate a secret key in the given situation because they have used the 
same basis, the names are also given in the last line.} 
\end{center}
\end{table}
As we have seen in the previous sections, if there would be no 
experimental errors, the protocol described in section 2 would, with 
minor modifications, become a quantum key distribution protocol between 
three parties. It would in principle
be enough for Trent to listen to Alice and Bob's public discussion
and perform the error correction and privacy amplification algorithm as
they do. To allow
Trent to share the final secret key with Alice and Bob, it is necessary
to modify the protocol, adding a modified error correction phase.

In the modified protocol we present in this section, 
the third party's name is Carol instead of Trent, who is only a
trusted arbitrator.
The difference between Trent and Carol is that Carol takes part in 
error correction and privacy amplification. 
Hence she actively takes part in the full protocol as a third member 
on equal footing with Alice and Bob. 

Concerning the error correction protocol, one possibility is to run 
it in two steps. First there 
is an error correction run by Alice and Carol, to which Bob listens and 
acts accordingly. At the end of this first run Carol and Alice keys are 
identical. Then Alice (or Carol) runs another error correction with Bob, 
now Carol (or Alice) listens and acts accordingly. 
Notice that Carol in this case must implement a full QKD node,
all phases of the protocol must be run by Carol, and at the end 
all three share the same identical key. 

In this case a more detailed analysis for the quantum channel is possible, 
in the sense that during the error correction part of the protocol, it 
will become evident where an error has occurred, if it was in channel 1 or 
channel 2, thanks to the fact that Carol participates in the error 
correction phase.

As it has been described in section 2, including an extra party means loosing 
more data in the sifting procedure. This is because the probability that 
the three parties all choose the same basis is only $1/4$. Which means 
that $3/4$ of the data remains un-used. However, it is actually possible 
to use a big 
part of the remaining data to create secret keys between Alice and Carol 
or Carol and Bob, because these parties in some cases use the same basis 
and hence the protocol reduces to the traditional BB84 protocol for only 
two parties. The different measurement combinations and who can create a 
secret key in a given situation is shown in table 1. In this way we find 
the following use of the data: $1/4$ of the data, columns 1 and 5, can be 
used to create a 
secret key between all three parties; $1/4$, cols.\ 2 and 6, for creating a 
secret key between Alice and Carol; $1/4$, cols.\ 4 and 8, for creating a 
secret key between  Carol and Bob; whereas the last $1/4$ of the data, 
cols.\ 3 and 7, has to be disregarded, since no secret key can be extracted 
from these data. 

This means that actually $3/4$ of the data can be used and only $1/4$ 
remains to be thrown away. Notice one interesting point namely that as 
long as data are distributed via a relay, Alice and Bob can not produce a 
secure {\it bipartite} key between them. This is because even if 
Trent/Carol is not actively participating in the error correction and 
privacy amplification part of the protocol, he/she  
will always have a lot of information on the key and has to be trusted.

In section 3 we argued that an eavesdropper could not benefit 
from eavesdropping in both channels, because with Trent as a trusted 
arbitrator in the middle, Alice and Bob effectively share one long quantum 
channel. However, this is no longer the case when Alice, Bob and Carol 
are equal partners, since in this case the three parties have and analyze 
independently the two channels. Moreover, for Eve to gain information on 
all three keys she obviously will be forced to eavesdrop in both channels.

\section{Relaying and extending the distance}
It would be very interesting to find ways to extend the reach of QKD 
without implementing entanglement swapping, teleportation 
or resorting to classical 
cryptography. At first sight it could seem that the protocol
presented in section 2 could {\sl always\/} 
extend the distance between Alice and Bob.
This is not true, and we recall here the reasons for this.

As of today, QKD is limited to a range of about 100km, the main limiting 
factor is due to losses. 
The main points of loss of photons in today QKD system using optical
fibers are (1) low efficiency of the detectors,  
approximately 10\%, (2) losses in the fibers, characterized by 
$\alpha$
in dB/Km which for the Telecom wavelength 1550nm has a typical value of
$0.25$dB/Km. The losses of the detector are independent of the 
distance between Alice and Bob, whereas the losses in the fibers
are the real limiting factor on the distance. 
In standard Telecom communications in optical fibers, these losses 
are compensated by repeaters, which in practice repopulate of the lost
photons the wave-packet. Since QKD uses single photons, once a
photon is absorbed by the fiber there could not be any kind of 
repeater which can recreate it. Thus quantum repeaters which work in a 
similar way as the classical ones do not exist. 

Let's consider Alice, Trent and Bob and assume that the transmission of
the line connecting Alice with Trent is $t_1$ and that of Trent with 
Bob is $t_2$,
and that the two lines have length $d_1$ and $d_2$ respectively.
Since when a photon is absorbed before reaching Trent, obviously Trent 
cannot send anything to Bob, Bob receives a fraction $t_1 t_2$ of the 
photons sent by Alice.

Consider now the case when Alice is directly connected to Bob with
a line of length $d_1 + d_2$. Since the transmission of a line is
given by
\begin{equation}
t = 10^{-\alpha d /10}
\end{equation}
the fraction of photons received by Bob is again $t_1 t_2$. 
This argument is of course very general 
and prevents the use of simple quantum relays to extend the maximum
distance of QKD systems in fibers.

The quantum relays we have presented in section 2 have thus mainly two
applications. The first is the possibility of realizing networks where
different platforms are used in different legs of the network, as optical 
fibers
and free space. The second is the possibility of realizing networks with
more than two participants, where the relay permits to connect many 
Alices to many Bobs on request. These will be discussed in the next section.

For completeness, in the Appendix we also describe a very simple classical 
protocol with QKD relays which can {\sl always\/} be used to extend 
the distance between Alice and Bob.

\section{Networks and Topology} 
The protocol we proposed in section 2 gives the possibility of
building more complicated network topologies. These in turn allow a
better use of resources and possibility of services used on-demand. 
Still there are some limitations, the principal one being the fact that
in the presence of $N$ relays, the average final secret key length
is reduced by at least a factor 
$1/2^{N+1}$ of the original qubits sent by Alice. 

The most suitable network that can be adopted with these relays seems to be 
a Star topology between each Trent and all his Alices and Bobs, and
a Full-mesh topology between all Trents. 
Thus for example one Trent
could cover a metropolitan area and be connected by optical fibers with
the Alices and Bobs, whereas the trust centers, i.e.\ different Trents, 
among themselves can be connected
with free space links, even through satellites.\footnote{Of course this
will be practically possible only if the transmission of the free space
link will be good enough to guarantee that enough qubits reach Bob, 
as we have discussed in section 5.}

\begin{figure}[t]
\begin{center}
\leavevmode
\hbox{%
\epsfxsize=4.5in
\epsffile{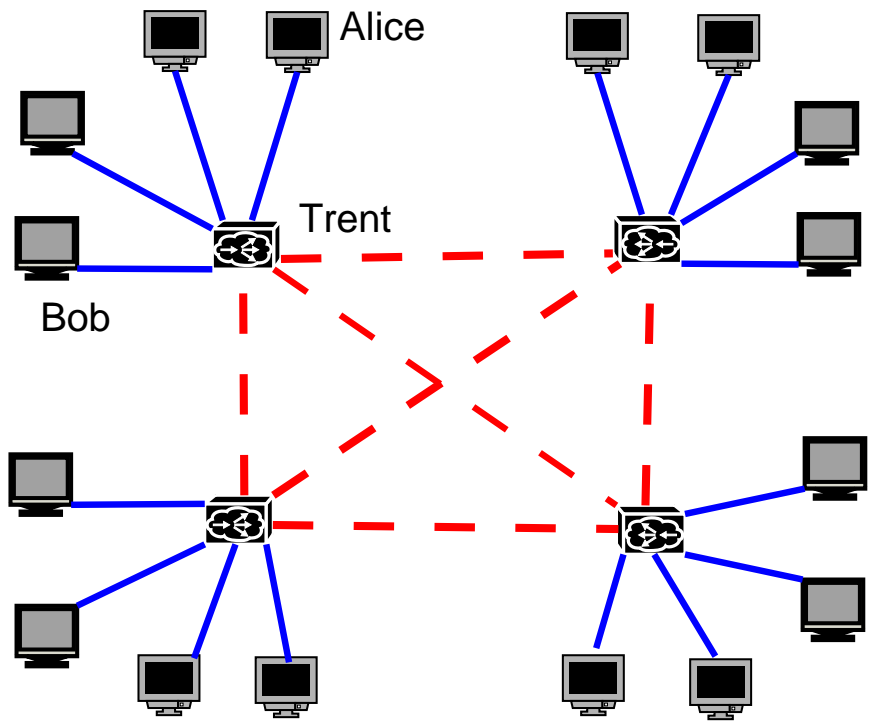}}
\caption{A simple network with Star topology between Alices and Bobs with
Trents as centers, and Full-Mesh between Trents}
\end{center}
\label{fig:fig3}
\end{figure}

In Figure 3 we give an
example of this network with four Trents. Notice that the number of links
between Trents grows as $N^2$, but that any link connecting one Alice to
one Bob will pass either one Trent, if Bob is in the same star as Alice,
or two Trents if Bob is in another star.

In these networks it is only possible to create secret keys
shared between one Alice and one Bob, and as we have seen in the previous
section, in case also with Trent/Carol.

\section{Conclusions}
We have introduced a quantum relay in order to form the basic leg in a quantum 
key distribution network. The quantum relay performs intercept/resend 
eavesdropping --- but collaborates with Alice and Bob in the sifting 
phase. The protocol can also be seen as a 
concatenation of quantum key 
distribution protocols. We have used the well-known BB84 protocol. The 
quantum relay, Trent, has to be trusted, since 
he in principle shares the same data with Alice and Bob.
The cost for introducing a quantum relay is a lower key generation rate. 
A minor modification in the protocol allows for multi user quantum key 
distribution.

One big advantage is that the proposed quantum relay can be implemented 
with todays technologies, 
and hence would make it possible to form small 
networks for quantum key distribution. Another advantage is that the 
nature of the relay is so simple that it allows to mix different quantum 
key distribution platforms. For example, one could imagine metropolitan 
areas connects via optical fiber links, whereas links between cities may 
be served by free space links. More results on networking and QKD
will be presented in ref.\ \cite{qcnet}.

It should be stressed again that the quantum relay has to be trusted,
which means that Trent collaborates fully with Alice and Bob. A first
analysis indicates that an eavesdropper who tries to learn the secret key
will not be able to get more information than in the standard BB84
protocol for two parties. And we conjecture that the security of the
protocol including Trent is the same as for the standard BB84.

\section*{Appendix}
In this Appendix we describe very shortly another protocol including
quantum relays similar to the previous one, but in which the relaying
part is completely classical. This will allow actually to obtain an
extension of the distance between Alice and Bob even with the current 
implementations in optical fibers.
Assume that Alice and Bob are connected
by $N$ relays linearly, as in a chain. The protocol is as follows:

\begin{enumerate}
\item Each relay runs a full BB84 (or any QKD) protocol with both
its right and left peer; Alice and Bob, who are at the ends of the chain,
run the protocol with the relay to their left/right; in this way $N+1$ 
different secret keys (possibly of different length) are established; each
relay knows 2 secret keys.
\item All relays announce publicy the XOR of the two secret keys (reducing
the longest key to the length of the shortest discarding the high order bits);
The role of the relays ends here.
\item Alice and Bob take the XOR of their secret key with the XOR-key announced
by the next relay thus discovering the other secret key of the relay; they
proceed in this way until they get the secret key of each other (in doing
these operations all keys are always reduced to the length of the shortest one
by discarding the high order bits).
\item Alice and Bob concatenate all secret keys and run a privacy amplification
algorithm on this; the order of magnitude of the shrinking parameter
can be estimated to be $NL$ bits where $L$ is the reduced length of each
secret key; The reason for this step is that in announcing publicy the 
XOR, Eve has learnt $NL$ bits of information which should be removed.
\end{enumerate}
The final secret key has length $L$, which in practice is of the order of the
shortest of the $N+1$ initial secret keys. With respect to the protocol 
discussed in this paper, this protocol does not shrink the final secret
key depending on the number of relays. On the other side the role of the
relays is much higher, since they have to run each {\it two full} QKD 
protocols, plus the managing the resulting secret keys. Moreover
in this case, also all relays can run steps 3 and 4, obtaining the 
exact final secret key.

\section*{Acknowledgment}
This work has been supported by EC under project SECOQC (contract n.
IST-2003-506813)


\end{document}